# Closed-Form Critical Conditions of Subharmonic Oscillations for Buck Converters

Chung-Chieh Fang

**Abstract**

A general critical condition of subharmonic oscillation in terms of the loop gain is derived. Many closed-form critical conditions for various control schemes in terms of converter parameters are also derived. Some previously known critical conditions become special cases in the generalized framework. Given an arbitrary control scheme, a systematic procedure is proposed to derive the critical condition for that control scheme. Different control schemes share similar forms of critical conditions. For example, both $V^2$ control and voltage mode control have the same form of critical condition. A peculiar phenomenon in average current mode control where subharmonic oscillation occurs in a window value of pole can be explained by the derived critical condition. A ripple amplitude index to predict subharmonic oscillation proposed in the past research has limited application and is shown invalid for a converter with a large pole.

## I. Introduction

Average *continuous-time* models are generally applied to analyze DC-DC converters [1]. The average model can predict some types of instabilities, but it generally cannot predict subharmonic oscillation [2] which is associated with a discrete-time pole at -1. By considering the sampling effects and increasing the system dimension, improved average models can predict the occurrence of subharmonic oscillation in some cases [3], [4]. However, even with positive gain or phase margins in the improved average model, the subharmonic oscillation may still occur [5]. This paper shows a systematic approach to accurately derive the critical condition.

Some subharmonic oscillation critical conditions for particular control schemes have been known. These control schemes are current mode control (CMC) [1] (where the subharmonic oscillation occurs at duty cycle $D = 1/2$), voltage mode control (VMC) without considering the effect of equivalent series resistance (ESR) [6], and $V^2$ control [7]. These conditions are generally obtained case by case, not systematically.

This paper tries to answer the following questions:

1) Is there a *general* closed-form critical condition that directly leads to these known conditions, and these known conditions become *special* cases in the generalized framework?

2) There exists a peculiar phenomenon in average current mode control such that subharmonic oscillation occurs in a *window* value of pole [8]. Is there a general theory to explain this phenomenon? Does it also exist in other cases?

3) It is hypothesized in [6] that the signal *ripple amplitude* can predict the occurrence of subharmonic oscillation. Does it have limited application only in some particular cases?

4) Some control schemes (such as $V^2$ control, VMC and CMC) may seem different, but do they share the same or similar *form* of critical condition?

5) Given an *arbitrary* control scheme, is there a *systematic* method to derive the critical condition for that control scheme?

The answers to all of these questions will be shown to be affirmative.

C.-C Fang is with Advanced Analog Technology, 2F, No. 17, Industry E. 2nd Rd., Hsinchu 300, Taiwan, Tel: +886-3-5633125 ext 3612, Email: fangcc3@yahoo.com

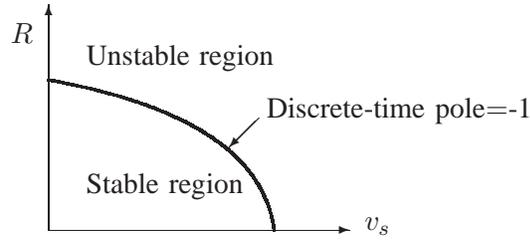

Figure 1. An illustrative critical boundary in the *parameter* space $(v_s, R)$.

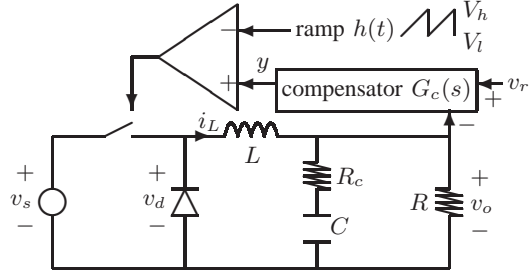

Figure 2. A VMC buck converter with a compensator $G_c(s)$.

Like the describing function approach [9], harmonic balance [10], [11] is a tool to analyze a nonlinear system. Based on harmonic balance, a critical condition of subharmonic oscillation in the buck converter is obtained in [12], [13], [14]. Since most converter designers are familiar with the loop gain analysis, here the critical condition in [12], [13], [14] is expressed in terms of the loop gain. Based on this critical condition, many closed-form critical conditions for various control schemes are obtained. To determine the loop gain for CMC, there exists different views about the PWM modulator gain [15]. For *stability* analysis, it will be shown that the PWM modulator gains for CMC and VMC are the same.

Note that, here, the critical condition is expressed in the converter *parameter space*. For example, given a source voltage $v_s$ and load resistance $R$, an illustrative critical boundary in the parameter space $(v_s, R)$ is shown in Fig. 1. The critical conditions defines the subharmonic oscillation boundary in the parameter space to separate stable and unstable regions. When a converter parameter crosses the critical value, the stability (or instability) changes. Critical conditions in *closed-forms* greatly facilitate the converter design, because the *quantitative* effect of each *relevant* converter parameter can be clearly seen.

The remainder of the paper is organized as follows. In Sections II and III, the operation of the buck converter and harmonic balance analysis are briefly reviewed. In Section IV, a general critical condition in terms of loop gain is derived. In Section V, the critical conditions for typical loop gains are derived. In Sections VI-IX, the proposed approach is systematically applied to various control schemes. Conclusions are collected in Section X.

## II. BRIEF REVIEW OF BUCK CONVERTER OPERATION

Consider a VMC buck converter shown in Fig. 2, where $v_s$ is the source voltage, $v_d$ is the voltage across the diode, $v_o$ is the output voltage, $v_r$ is the reference voltage, $y$ is the compensator output signal, and $G_c(s) = -y(s)/v_o(s)$ is the compensator transfer function. Denote the ramp slope as $m_a = \dot{h}(d)$. Denote the cycle period as $T$, the switching frequency as $f_s = 1/T$ and let $\omega_s := 2\pi f_s$.

Similarly, a CMC buck converter is shown in Fig. 3, where a control signal $i_c$ controls the (peak) inductor current $i_L$, and $y = i_c - i_L$ (equivalently, $G_c(s) = 1$). Note that in Fig. 3, the circuit is arranged in a way to have a similar output signal $y$ as in VMC.

The operation of the converter is as follows. Within a cycle period $T$, the dynamics is switched between two stages, $S_1$ and $S_2$. Switching occurs when the compensator output $y$ intersects with the ramp signal $h(t) := V_l + (V_h - V_l)(\frac{t}{T} \bmod 1)$, where $h(t)$ varies from a low value $V_l$ to a high value $V_h$. Denote the ramp amplitude as $V_m = V_h - V_l$. In the trailing-edge modulation (TEM) [2], $v_d = v_s$ in stage $S_1$ and $v_d = 0$ in stage $S_2$. The waveform of $v_d(t)$ is a square wave with a duty cycle $D$. In the leading-edge modulation (LEM), $v_d = 0$ in stage $S_1$ and $v_d = v_s$ in stage $S_2$. *This paper focuses on TEM only. For LEM, see [12], [13], [14].* The controlled buck converter is equivalent to a nonlinear switching model shown in Fig. 4. Note that $v_r$ does not affect the loop stability, and it adjusts the DC offset of $h(t)$ in Fig. 4. Let the loop gain be $T(s)$. From Fig. 4, $T(s) = v_s G(s)/V_m$, and the model in Fig. 4 can be normalized as in Fig. 5.

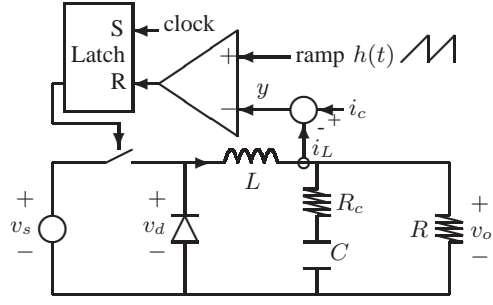

Figure 3. A CMC buck converter.

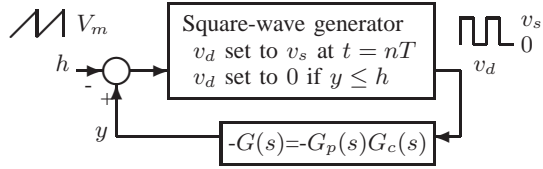

Figure 4. An equivalent nonlinear switching model.

## III. Brief Review of Harmonic Balance Analysis

A brief review of harmonic balance analysis based on [12], [13], [14] is presented. The square-wave $v_d(t)$ can be represented by Fourier series (harmonics). The signal path in the control *loop* has two parts: from $y$ to $v_d$ through a *nonlinear* PWM modulator, and from $v_d$ to $y$ through a *linear* transfer function $G(s) := G_p(s)G_c(s)$, where $G_p(s)$ is the power stage transfer function and its representation depends on what signal is fed to the compensator.

In VMC, $v_o$ is fed to the compensator. Let $\rho = R/(R + R_c)$. For $R_c = 0$, $\rho = 1$. From [1, p. 470], the power stage $v_d$-to-$v_o$ transfer function is

$$G_p(s) = \frac{sR_cC + 1}{\frac{LCs^2}{\rho} + (\frac{L}{R} + R_cC)s + 1} \tag{1}$$

In CMC, $i_L$ is fed to the compensator (with $G_c(s) = 1$). From [1, p. 470], the power stage $v_d$-to-$i_L$ transfer function is

$$G_p(s) = \frac{\frac{Cs}{\rho} + \frac{1}{R}}{\frac{LCs^2}{\rho} + (\frac{L}{R} + R_cC)s + 1} \tag{2}$$

Let $x^0(t)$ be the $T$-periodic solution of the converter. Let $y^0(t)$ be the corresponding $T$-periodic compensator output signal. The intersection of $h(t)$ with $y^0(t)$ determines the duty cycle and hence the waveform of $v_d(t)$. By "balancing" the equation $y^0(t) = h(t)$ (written in Fourier series form) at the switching instants, conditions for existence of periodic solutions and subharmonic oscillation can be derived. Let $\mathbf{Re}$ denote taking the real part of a complex number. Based on [12], [13], [14], the subharmonic oscillation occurs when

$$2v_s\mathbf{Re}\left[\sum_{k=1}^{\infty}[(1 - e^{j2k\pi D})G(jk\omega_s) - G(j(k - \frac{1}{2})\omega_s)]\right] = V_m \tag{3}$$

This critical condition is valid for a *general* switching system shown in Fig. 4. It will be shown later that this critical condition can be expressed in a closed-form related to the hyperbolic function $\mathrm{csch}$.

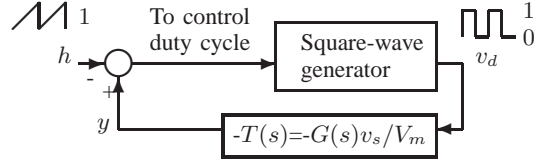

Figure 5. An equivalent *normalized* nonlinear switching model.

## IV. CRITICAL CONDITION IN TERMS OF LOOP GAIN

Since $T(s) = v_s G(s)/V_m$, the critical condition (3) directly leads to the following result.

*Consider a closed-loop buck converter with a loop gain $T(s)$. For $V_m \neq 0$, the critical condition of subharmonic oscillation is*

$$2\mathbf{Re}\left[\sum_{k=1}^{\infty}(1-e^{j2k\pi D})T(jk\omega_s) - T(j(k-\frac{1}{2})\omega_s)\right] = 1 \tag{4}$$

Note that (4) is an expression of convenience. For $V_m = 0$, the loop gain $v_s G(s)/V_m$ would be infinite. In that case, the equivalent critical condition (3) is used. Also note that (4) is valid for *both* VMC and CMC by assuming the *same* PWM modulator gain to be $1/V_m$. Since both (3) and (4) are exact critical conditions, they can be used as benchmarks to determine the accuracy of other critical conditions.

Generally, the power stage has an order of at least two (associated with $L$ and $C$). If the compensator has an order of three, such as the type-III compensator [16], the loop gain $T(s)$ has an order of five. However, there are ways to simplify the analysis without losing the accuracy.

First, the compensator is generally designed to cancel some poles or zeros of the power stage, and the order of $T(s)$ is reduced. Second, from (4), one sees that the shape of the Bode plot for the frequency smaller than $\omega_s/2$ is irrelevant. One can use a simplified high-frequency form of $T(s)$ for stability analysis. Third, by decomposing the loop gain into *partial fractions*, many new *closed-form* critical conditions in terms of the converter parameters can be obtained.

Define an "F-transform" of $T(s)$ as

$$\mathcal{F}[T(s)] := 2\mathbf{Re}\left[\sum_{k=1}^{\infty}(1-e^{j2k\pi D})T(jk\omega_s) - T(j(k-\frac{1}{2})\omega_s)\right] \tag{5}$$

Note that $\mathcal{F}[T(s)]$ is a function of many converter parameters. Here, this function is called an L-plot, denoted as $\mathcal{L}$. For example, let it be a function of $D$ and it becomes $\mathcal{L}(D)$. Then, the critical condition (4) becomes $\mathcal{L} := \mathcal{F}[T(s)] = 1$. The critical condition itself does not tell which side of the critical boundary will be the stable region. Generally for a converter, the region with $\mathcal{L} < 1$ is stable. The F-transforms of typical loop gain functions are presented next.

## V. "F-TRANSFORMS" OF TYPICAL LOOP GAIN FUNCTIONS

The loop gain is generally designed to have sufficient gain and phase margins. For a stable converter, the phase of $T(j\omega)$ is less than $180°$ at the crossover frequency $\omega_c$. One can focus only those loop gains of first or second orders.

The loop gain can be further decomposed into a combination of partial fractions. Only partial fractions of first orders are considered. Similar analysis can be applied to partial fractions of second orders.

Let $\omega_p$ and $\omega_z$ be the pole and zero of $T(s)$. Let $p = \omega_p/\omega_s$ and $z = \omega_z/\omega_s$. The F-transform of the fraction $1/(s + \omega_p)$ will be the building block to derive the F-transforms of other loop gains.

Table I
F-TRANSFORM OF TYPICAL LOOP GAIN $T(s)$.

| Case | $T(s)$ | $\mathcal{F}[T(s)]$ | (note: $p = \omega_p/\omega_s$ and $z = \omega_z/\omega_s$) |
|------|--------|---------------------|---|
| $\mathcal{C}_1$ | $\frac{1}{s+\omega_p}$ | $\frac{1}{\omega_s}\alpha(D,p) = \frac{1}{\omega_s}(\alpha_0(D) - \alpha_1(D)p + c(D,p))$ | |
| $\mathcal{C}_2$ | $\frac{1}{s}$ | $\frac{1}{\omega_s}\alpha_0(D)$ | |
| $\mathcal{C}_3$ | $\frac{1}{1+s/\omega_p}$ | $p\alpha(D,p)$ | |
| $\mathcal{C}_4$ | $\frac{1+s/\omega_z}{1+s/\omega_p}$ | $-\frac{p}{z} + p(1-\frac{p}{z})\alpha(D,p)$ | |
| $\mathcal{C}_5$ | $\frac{1}{s(1+s/\omega_p)}$ | $\frac{1}{\omega_s}(\alpha_1(D)p - c(D,p)) = \frac{1}{\omega_s}(\alpha_0(D) - \alpha(D,p))$ | |
| $\mathcal{C}_6$ | $\frac{1}{s^2}$ | $\frac{1}{\omega_s^2}\alpha_1(D)$ | |
| $\mathcal{C}_7$ | $\frac{1+s/\omega_z}{s^2}$ | $\frac{1}{\omega_s^2}(\frac{1}{z}\alpha_0(D) + \alpha_1(D))$ | |
| $\mathcal{C}_8$ | $\frac{1+s/\omega_z}{s(1+s/\omega_p)}$ | $\frac{1}{\omega_s}(\frac{p}{z}\alpha_0(D) - (\frac{p}{z}-1)(\alpha_1(D)p - c(D,p)))$ | |
| $\mathcal{C}_9$ | $\frac{1+s/\omega_z}{s^2(1+s/\omega_p)}$ | $\frac{1}{\omega_s^2}(\frac{p}{z}\alpha_1(D) + (\frac{1}{p}-\frac{1}{z})c(D,p))$ | |

From (5),

$$
\omega_s\mathcal{F}[\frac{1}{s+\omega_p}]
$$
$$
= 2\mathbf{Re}\left[\sum_{k=1}^{\infty}(1-e^{j2k\pi D})\frac{1}{jk+p} - \frac{1}{j(k-1/2)+p}\right]
$$
$$
= 2\pi\operatorname{csch}(2\pi p) - \pi e^{\pi p(1-2D)}\operatorname{csch}(\pi p) \qquad (6)
$$
$$
:= \alpha(D,p)
$$

where the proof of (6) can be obtained by looking up from a handbook of mathematical formulas or checked by a simple computer program, and $\operatorname{csch}$ is a hyperbolic function. Using Taylor series expansion, let $\alpha(D,p) = \sum_{k=0}^{\infty}(-1)^k\alpha_k(D)p^k$. Using the L'Hospital's rule, one has $\alpha_0(D) = \pi(2D-1)$ and $\alpha_1(D) = \pi^2(2D^2 - 2D + 1)$. A plot of $\alpha(D,p)$ is shown in Fig. 6. One sees that $\alpha(D,p)$ is close to zero for $D > 0.5$ and $p > 1$. The right straight-line edge in Fig. 6 is $\alpha(D,0) = \alpha_0(D) = \pi(2D-1)$.

Let the correction term be $c(D,p) = \sum_{k=2}^{\infty}(-1)^k\alpha_k(D)p^k$. One has

$$
c(D,p) = \alpha(D,p) - \alpha_0(D) + \alpha_1(D)p \qquad (7)
$$

A plot of $c(D,p)$ is shown in Fig. 7. For a large $p$, since $\alpha(D,p)$ is small, $c(D,p)$ increases almost linearly as a function of $p$ shown in Fig. 7. From Fig. 7, the correction term $c(D,p)$ is significant only if $p > 0.1$ (equivalently, $\omega_p > 0.1\omega_s$). For $p < 0.1$, $c(D,p)$ can be ignored.

The F-transforms of other loop gains are shown in Table I. For each case $\mathcal{C}_1$-$\mathcal{C}_9$, the F-transform can be derived or proved (by simple algebra) in three ways. First, follow the definition of the F-transform as in (5). Second, by decomposing $T(s)$ into a combination of fractions $1/s$, $1/(s+\omega_p)$, etc, $\mathcal{F}[T(s)]$ is a combination of $\mathcal{F}[1/s]$, $\mathcal{F}[1/(s+\omega_p)]$, etc. For example,

- Use of $\mathcal{C}_1$ and $\mathcal{C}_2$ leads to $\mathcal{C}_5$ and $\mathcal{C}_8$;
- Use of $\mathcal{C}_2$ and $\mathcal{C}_6$ leads to $\mathcal{C}_7$;
- Use of $\mathcal{C}_1$, $\mathcal{C}_2$ and $\mathcal{C}_6$ leads to $\mathcal{C}_9$; and
- Use of $\mathcal{C}_1$ and the fact that $\mathcal{F}[1] = -1$ leads to $\mathcal{C}_4$.

Third, each case is a special/general case of other cases. For example, by setting $\omega_p \to 0$, $\mathcal{C}_1$ leads to $\mathcal{C}_2$; $\mathcal{C}_5$ leads to $\mathcal{C}_6$; and $\mathcal{C}_8$ leads to $\mathcal{C}_7$. By setting $\omega_z \to \infty$, $\mathcal{C}_7$ leads to $\mathcal{C}_6$. One sees that $\mathcal{C}_1$ is a building block for other cases because they have *similar terms* as $\mathcal{C}_1$.

*Remarks:*

(a) All of the transforms in Table I are *exact*. No approximation is assumed.

(b) There is no correction term $c(D,p)$ for $\mathcal{C}_2$, $\mathcal{C}_6$, or $\mathcal{C}_7$. All other cases have a correction term $c(D,p)$, which is small and can be ignored if $p < 0.1$ as discussed above.

The critical conditions for various control schemes are readily derived next.

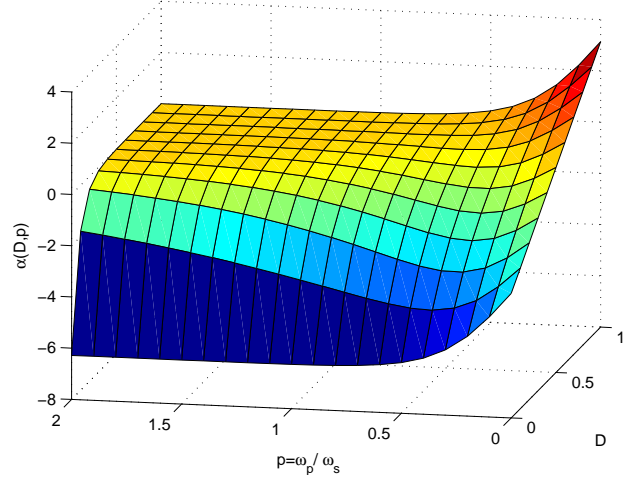

Figure 6. Plot of $\alpha(D, p)$ for case $\mathcal{C}_1$.

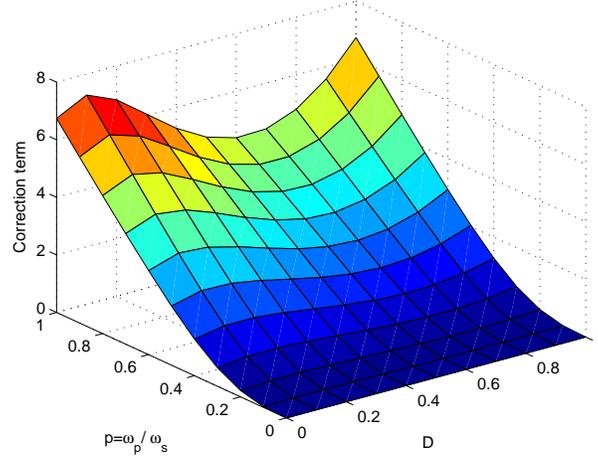

Figure 7. Plot of the correction term $c(D, p)$.

## VI. CMC: CASE $\mathcal{C}_2$

From Fig. 3, $T(s) = G_p(s)v_s/V_m$. From (2), at high frequency, $G_p(s) \approx 1/Ls$. One has $T(s) \approx v_s/LV_m s$, which is of case $\mathcal{C}_2$ in Table I. Its F-transform leads to the well-known critical condition for CMC [1], [5]

$$\frac{v_s \alpha_0(D)}{L V_m \omega_s} = 1 \tag{8}$$

or equivalently,

$$\frac{v_s}{L}\left(D - \frac{1}{2}\right) = \frac{V_m}{T} = m_a \tag{9}$$

Without the compensation ramp ($m_a = 0$), the critical point is $D = 1/2$.

## VII. PROPORTIONAL VOLTAGE MODE CONTROL (PVMC): CASES $\mathcal{C}_7$, $\mathcal{C}_6$, $\mathcal{C}_5$, AND $\mathcal{C}_3$

In proportional voltage mode control (PVMC), let the voltage loop have a proportional feedback gain $k_p$. One has $G_c(s) = k_p$ and $y = k_p(v_r - v_o)$. Five cases are considered to see the effects of $R_c$ and $R$ on the subharmonic oscillation.

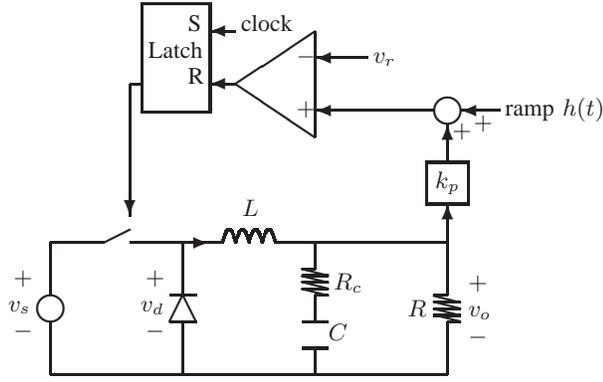

Figure 8. A buck converter under CF-PVR.

## A. PVMC with $R_c \neq 0$: Case $\mathcal{C}_7$

From (1), at high frequency, $G_p(s) \approx \rho(sR_cC + 1)/V_mLCs^2$, and

$$T(s) = \frac{v_s G_c(s) G_p(s)}{V_m} \approx \frac{v_s k_p \rho}{V_m LC}\left(\frac{1 + s/\omega_z}{s^2}\right) \tag{10}$$

which is of case $\mathcal{C}_7$ with $\omega_z = 1/R_cC$. From Table I, the critical condition is

$$\frac{v_s k_p \rho T^2}{4V_m LC}\left[\frac{2R_cC}{T}(2D - 1) + (2D^2 - 2D + 1)\right] = 1 \tag{11}$$

which is a weighted combination of $2D - 1$ (a condition for CMC as seen in (9)) and $2D^2 - 2D + 1$. The term $2D - 1$ can be ignored only if $R_cC \ll T$. Therefore, in PVMC with a large $R_c$, the critical condition has a term $2D - 1$ like CMC.

## B. $V^2$ Control: Case $\mathcal{C}_7$, Same as PVMC

The constant-frequency peak voltage regulator (CF-PVR, as shown in Fig. 8), a type of $V^2$ control [7], is proved below to be a special case of PVMC in terms of the critical condition. In CF-PVR, the output voltage is sensed (through a voltage divider with a gain of $k_p$), added with a stabilization ramp $h(t)$, and compared with a reference signal $v_r$ to determine the duty cycle. One has $G_c(s) = k_p$, same as PVMC. Therefore, they have the same form of critical condition (11).

For the buck converter, $v_o = Dv_s$. To avoid subharmonic oscillation, (11) is rearranged as (inequality)

$$\frac{m_a L}{k_p \rho v_o R_c} > \frac{2D - 1}{2D} + \frac{T}{R_cC}\left(\frac{1 - 2D}{4D} + \frac{D}{2}\right) \tag{12}$$

which agrees with [7, Eq. 7] and shows the required ramp slope $m_a$ to stabilize the subharmonic oscillation.

Note that the critical condition (12) has two terms. The second term is related to the output capacitor $C$. For very large $C$, the second term can be ignored, and the critical condition (12) is rearranged as (with $k_p = 1$ and $\rho \approx 1$)

$$m_a > \frac{2D - 1}{2}\left(\frac{v_s R_c}{L}\right) = \frac{R_c(m_1 - m_2)}{2} \tag{13}$$

where $m_1$ and $m_2$ are the inductor current slopes in the stages $S_1$ and $S_2$ respectively. Note that (13) shows the required ramp slope $m_a$ to stabilize the subharmonic oscillation when the output capacitor is very large. The critical condition (13) is reasonable, because, with very large output capacitor, the output voltage ripple is mostly contributed by the inductor current (multiplied by $R_c$) and the straight-line inductor current analysis (as in the current-mode control) is adequate for stability analysis. When the output capacitor is small, the critical condition (12) is more accurate, with an additional (second) term related to the output capacitor ripple.

Without the ramp ($m_a = 0$), rearranging (12), subharmonic oscillation is avoided if

$$\frac{T}{R_cC} < \frac{1}{\frac{1}{2} + \frac{D^2}{1 - 2D}} \tag{14}$$

or, equivalently,

$$\frac{R_cC}{T} > \frac{1}{2} + \frac{D^2}{1-2D} \text{ and } D < \frac{1}{2} \tag{15}$$

also agreed with [7, Eq. 5].

*Remarks:*

(a) The conditions (12) is applicable to *both* PVMC and CF-PVR, and (14) or (15) is a special case ($m_a = 0$) of (12).

(b) $D < 1/2$ is explicitly required in (15), whereas $D < 1/2$ is implicitly required in (14).

### C. PVMC with $R_c = 0$: Case $\mathcal{C}_6$

Fro $R_c = 0$, the zero $\omega_z$ is at infinity and $\mathcal{C}_7$ becomes $\mathcal{C}_6$. Either from Table I or (11), the critical condition is

$$\frac{v_s k_p T^2}{4 V_m L C}(2D^2 - 2D + 1) = 1 \tag{16}$$

agreed with [6] which claims that the ripple amplitude such as $\Delta v_o$ (or $\Delta y$) is an index to predict the subharmonic oscillation. In the buck converter, it *happens* that $\Delta v_o = (D^2 - D)T^2 v_s/8LC$ [1], and the critical condition (16) can be expressed as a condition in terms of $\Delta v_o$. However, the condition (16) is valid only for $R_c = 0$.

### D. PVMC with $R_c = 0$ and Small $R$: Case $\mathcal{C}_5$

As the load resistance $R$ decreases, a pole at $\omega_p = 1/RC$ becomes significant. At high frequency, from (1),

$$T(s) \approx (\frac{v_s k_p R}{V_m L})\frac{1}{s(1+s/\omega_p)} \tag{17}$$

which is of case $\mathcal{C}_5$. The critical condition expressed in terms of $v_s$ is

$$v_s = \frac{L V_m \omega_s}{R k_p(\alpha_0(D) - \alpha(D, p))} \tag{18}$$

One sees that as $R$ decreases, the critical value of $v_s$ increases proportionally according to $1/R(\alpha_0(D) - \alpha(D, p))$.

### E. PVMC with Simple $RL$ Circuit: Case $\mathcal{C}_3$

Most reported subharmonic oscillations occur in switched $RLC$ circuits. However, subharmonic oscillation may occur even in a simple switched $RL$ circuit of *first* order.

**Example 1.** (*Subharmonic oscillation occurs even with phase margin of* $90.6°$ *and infinite gain margin in the average model.*) Consider a simple $RL$ circuit shown in Fig. 9. It is actually equivalent to peak current (voltage) control. From Fig. 6, for *small* $p$ and $D > 1/2$, subharmonic oscillation may occur.

Let the converter parameters be $v_s = 10$ V, $v_r = 7.5$, $V_l = 0$, $V_h = 1$, $f_s = 1$ MHz, $L = 1$ $\mu$H, and $R = 1$ $\Omega$. One has $p = \omega_p/\omega_s = R/L\omega_s$ and

$$T(s) = \frac{v_s k_p}{V_m(1 + Ls/R)} \tag{19}$$

For $k_p = 9$, the Bode plot (Fig. 10) shows a phase margin of $90.6°$ and an infinite gain margin. However, subharmonic oscillation still occurs as shown next.

Based on the *exact* sampled-data model [12], the pole as a function of $k_p$ is shown in Fig. 11, which shows the occurrence of subharmonic oscillation at $k_p = 8.63$. Note that (19) is of case $\mathcal{C}_3$ or $\mathcal{C}_1$. From Table I, the critical condition is

$$v_s k_p p \alpha(D, p) = V_m \tag{20}$$

Solving (20) and the steady-state condition $k_p(v_r - v_o) = h(d)$, the critical gain is $k_p = 8.52$, close to the exact sampled-data analysis.

First, let $k_p = 8$. The $T$-periodic orbit is stable as shown in Fig. 12.

Next, let $k_p = 9$. Subharmonic oscillation occurs with a $2T$-periodic orbit as shown in Fig. 13. $\qquad\square$

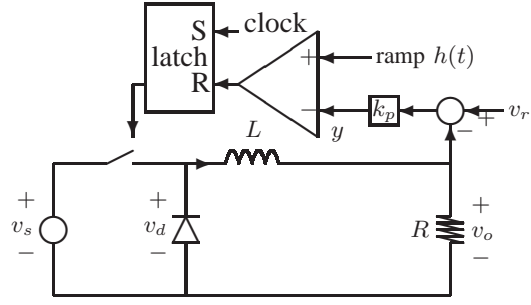

Figure 9.   Simple PVMC $RL$ circuit.

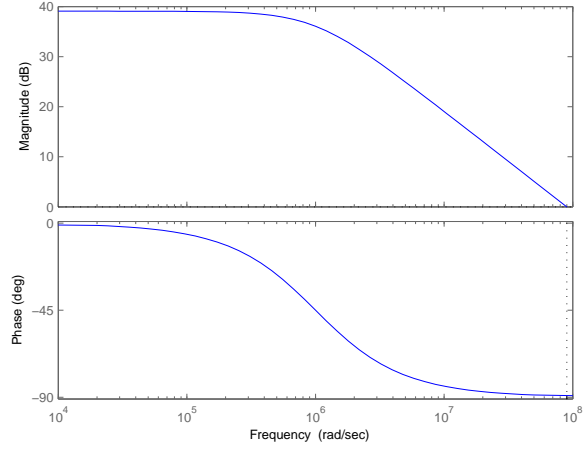

Figure 10.   Bode plot of the $RL$ circuit shows a phase margin of $90.6°$ and an infinite gain margin.

## VIII. Average Current Mode Control (ACMC) with a Type-II Compensator: Case $\mathcal{C}_5$

The operation of ACMC is as follows [16]: The inductor current $i_L$ is sensed by a resistor $R_s$ and compared with a voltage reference $v_r$ from the voltage loop. The difference is amplified by a current-loop compensator, generally a type-II compensator,

$$G_c(s) = \frac{R_s K_c (1 + s/z_c)}{s(1 + s/\omega_p)} \tag{21}$$

which has a small zero $z_c \ll \omega_s$, an integrator pole 0, a large pole $\omega_p$, and a gain $K_c$. At high frequency, from (2),

$$T(s) = \frac{v_s G_c(s) G_p(s)}{V_m} \approx \left(\frac{v_s R_s K_c}{V_m z_c L}\right) \frac{1}{s(1 + s/\omega_p)} \tag{22}$$

which is of case $\mathcal{C}_5$. The critical condition is

$$\frac{v_s R_s K_c}{V_m z_c L \omega_s}(\alpha_0(D) - \alpha(D, p)) = 1 \tag{23}$$

It can be also expressed in terms of the ramp slope. From (23), the minimum ramp slope to avoid the subharmonic oscillation is

$$m_a = \frac{V_m}{T} > \frac{v_s R_s K_c}{2\pi z_c L}(\alpha_0(D) - \alpha(D, p)) \tag{24}$$

A plot of $\alpha_0(D) - \alpha(D, p)$ (also a scaled L-plot $\mathcal{L}(D, p)$) is shown in Fig. 14. The corresponding contour plot is shown in Fig. 15.

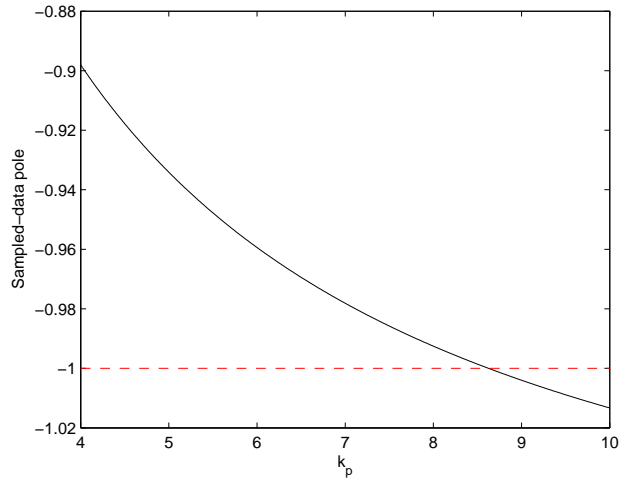

Figure 11. Sampled-data pole as a function of $k_p$. Subharmonic oscillation occurs at $k_p = 8.64$.

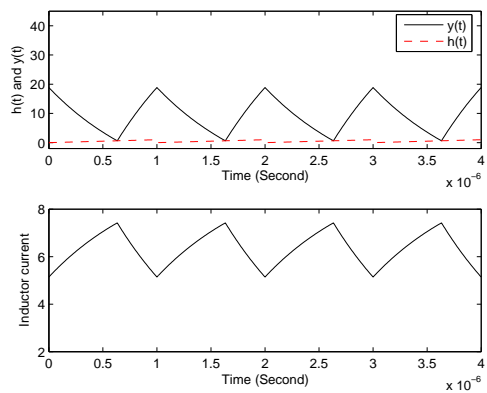

Figure 12. Stable $T$-periodic orbit, $k_p = 8$.

The contour plot is helpful to determine the maximum allowable loop gain without subharmonic oscillation. For example, suppose the ACMC converter is designed to operate at $D = 0.7$ and $p = 0.4$. From the contour plot, the L-plot value $(\alpha_0(D) - \alpha(D, p))$ is around 1.5 (also the "elevation" or "contour level" in the contour plot). Let $K = v_s R_s K_c / V_m z_c L \omega_s$. Then, from (23), $K < 2/3$ is required to avoid the subharmonic oscillation. Since the largest value of $\alpha_0(D) - \alpha(D, p)$ is $\pi$, the subharmonic oscillation is completely avoided for any $D$ or $p$ if $K < 1/\pi$.

The contour plot also defines the critical boundary. For example, if $K = 1$, the contour plot with the elevation at 1 shown in Fig. 15 defines the critical boundary which separate stable region from the unstable region in the $(D, p)$ space. Only those regions in the $(D, p)$ space with the elevation less than 1 are the stable operating regions.

The contour plot is also helpful to see the "weak spots" which are susceptible to subharmonic oscillation in the parameter space. Those regions with high elevations in the contour plot are the weak spots. In Fig. 15, one sees that for a large or small $D$, the elevation is high, and these regions are the weak spots. For mid-range of $D$, there exists a *window* of $p$ such that the elevation is high. For example, in Fig. 15, draw a line at $D = 0.6$. As $p$ increases, for $p \in (0.2, 0.5)$, the plot rises above 1 and then falls below 1. In this window of $p$, the converter is prone to subharmonic oscillation.

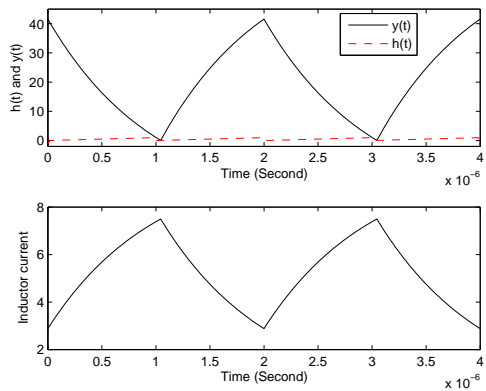

Figure 13. Subharmonic oscillation with $2T$-periodic orbit, $k_p = 9$.

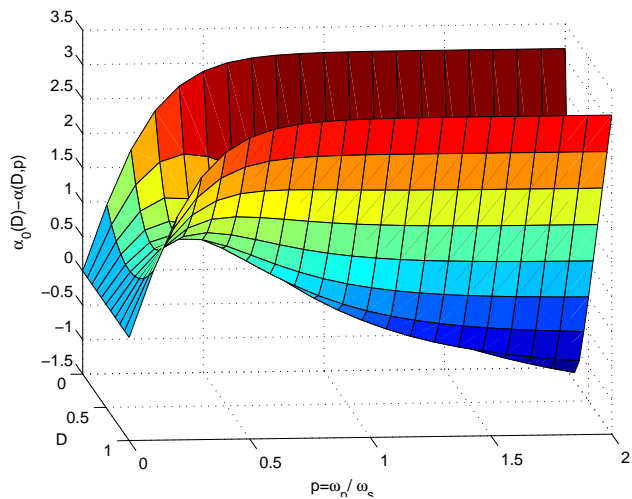

Figure 14. Plot of $\alpha_0(D) - \alpha(D, p)$ for case $\mathcal{C}_5$.

The instability window of $p$ can be estimated. Since

$$
\alpha_0(D) - \alpha(D, p) \approx
\begin{cases}
\alpha_1(D)p & \text{(for a small } p\text{)} \\
\alpha_0(D) - 4\pi e^{-2\pi p} + 2\pi e^{-2\pi Dp} & \text{(for a large } p\text{)}
\end{cases}
\tag{25}
$$

Based on (23) and (25), the instability window of $p$ is

$$
\left( \frac{1}{K\alpha_1(D)}, \frac{1}{2} + \frac{2D - 1 + 2e^{-\pi D} - \frac{1}{K\pi}}{4\pi D e^{-\pi D}} \right)
\tag{26}
$$

From (23) and Fig. 15, for $p$ being inside this window, with $K \approx 1$ and $0.1 < D < 0.7$, the subharmonic oscillation would occur, as shown in the next example.

**Example 2.** (*Instability window of $p$.*) Consider a buck converter under ACMC [17, p. 114]. The power stage parameters are $v_s = 14$ V, $v_o = 5$ V, $v_r = 0.5$, $V_l = 0$, $V_h = 1$, $f_s = 50$ kHz, $L = 46.1$ $\mu$H, $C = 380$ $\mu$F with ESR $R_c = 0.02$ $\Omega$, and $R = 1$ $\Omega$. The inductor current sensing resistance $R_s$ is 0.1 $\Omega$. The compensator has a zero at $z_c = 5652.9$ rad/s, poles at 0 and $\omega_p$, and a gain $K_c = 75506$.

**Time-domain simulation.** The compensator pole $\omega_p$ is varied from $0.14\omega_s$ to $0.81\omega_s$. An unstable *window* of $\omega_p$ between $0.18\omega_s$ and $0.49\omega_s$ was found and reported in [8]. When $\omega_p$ is inside the window, the subharmonic oscillation occurs.

Figure 15. Contour plot of $\alpha_0(D) - \alpha(D, p)$ for case $\mathcal{C}_5$.

Figure 16. Stable $T$-periodic solution, $\omega_p = 0.15\omega_s$.

For example, let $\omega_p = 0.15\omega_s$. The converter is stable (Fig. 16). Let $\omega_p = 0.49\omega_s$. The converter is unstable with subharmonic oscillation (Fig. 17). Next, let $\omega_p = 0.81\omega_s$. The converter is stable again (Fig. 18).

**Independent sampled-data analysis.** The sampled-data pole trajectories for $\omega_p/\omega_s \in (0.1, 0.8)$ are shown in Fig. 19. There are four poles. Two poles are fixed around 0.88, and 0.95 ($\approx e^{\frac{-T}{RC}}$). A pole leaves the unit circle through -1 when $\omega_p = 0.18\omega_s$, and enters the unit circle when $\omega_p = 0.49\omega_s$. This explains exactly the instability window of $\omega_p$.

**Accurate prediction by the L-plot.** The L-plot from (23), as a function of $p$ for $D = v_o/v_s = 0.357$ is shown in Fig. 20. It shows the instability window of $p \in [0.18, 0.46]$. The small error is due to the approximation of $T(s)$ in (22). The L-plot is also a *scaled* cross-section at D=0.357 in Fig. 14. The instability window of $p$ predicted by (26) is $[0.15, 0.58]$, which is also a close estimation.

In ACMC, the subharmonic oscillation is unrelated to the ripple *amplitude* [8]. As $\omega_p$ increases, the amplitude of $y^0(t)$ increases monotonously (because $G_c(s)$ is a low-pass filter). For $\omega_p = 0.49\omega_s$, the converter is unstable with a small amplitude of $y^0(t)$, whereas for $\omega_p = 0.81\omega_s$, the converter is stable with a larger amplitude of $y^0(t)$. The ripple *amplitude* index to predict the subharmonic oscillation hypothesized in [6] does not apply in this case.

In this example, the existence of instability window of $\omega_p$ is verified by time-domain simulation, pole trajectories, and agreed closely with the derived critical condition. □

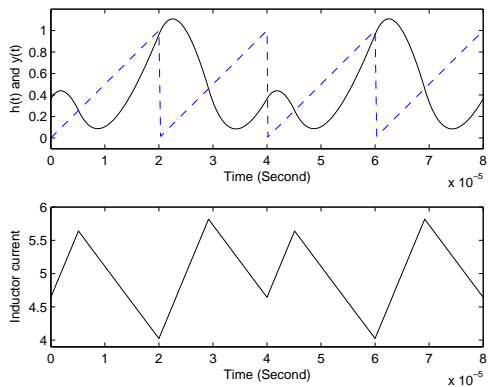

Figure 17. Subharmonic oscillation, $\omega_p = 0.49\omega_s$.

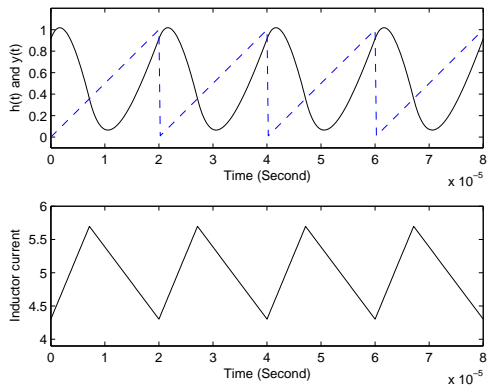

Figure 18. Stable $T$-periodic solution, $\omega_p = 0.81\omega_s$.

## IX. VMC with a Type-III Compensator: Case $\mathcal{C}_5$

A typical guideline [16, p. 412] popular in industry to set the parameters of the type-III three-pole-two-zero compensator is as follows. Set one pole at 0 as an integrator, the second pole at $1/R_cC$ (to cancel the power stage zero), and the third pole at $\omega_p = \omega_s/2$. Set the gain $K_c$ to adjust the phase margin and the crossover frequency. Set the two zeros at $\kappa_z/\sqrt{LC}$ and $1/\sqrt{LC}$ (to cancel the power stage poles), where $\kappa_z$ is a zero scale factor to have additional flexibility to adjust the phase margin and the crossover frequency. The zero scale factor $\kappa_z$ used in industry typically varies between 0.1 and 1.2. As will be shown later, a smaller value of $\kappa_z$ may lead to the subharmonic oscillation. Taking into account the above guidelines, the compensator has a transfer function

$$G_c(s) = \frac{K_c(1 + \sqrt{LC}s/\kappa_z)(1 + \sqrt{LC}s)}{s(1 + s/\omega_p)(1 + R_cCs)} \tag{27}$$

From (1), the loop gain is

$$T(s) = \frac{v_s G_c(s) G_p(s)}{V_m} \tag{28}$$

$$\approx \frac{v_s K_c \rho}{V_m \kappa_z} \frac{1}{s(1 + s/\omega_p)} \quad \text{(at high frequency)} \tag{29}$$

which is of case $\mathcal{C}_5$. The critical condition expressed in terms of $v_s$ is

$$v_s = \frac{V_m \kappa_z \omega_s}{K_c \rho(\alpha_0(D) - \alpha(D, p))} \tag{30}$$

As discussed above, a smaller $\kappa_z$ leads to a smaller critical source voltage $v_s$.

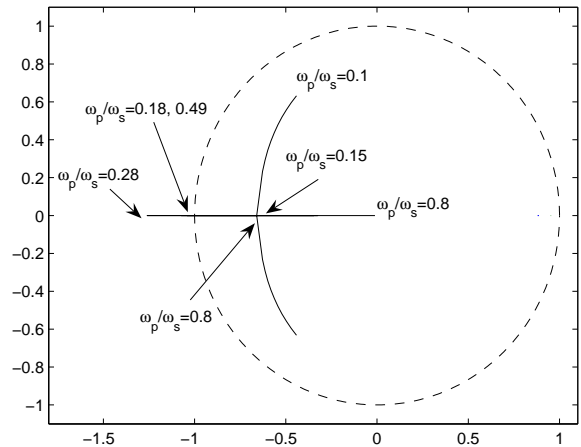

Figure 19. Sampled-data pole trajectories for $\omega_p/\omega_s \in (0.1, 0.8)$.

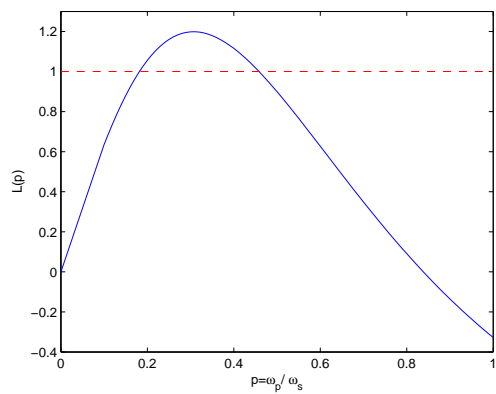

Figure 20. L-plot (23) as a function of $p$. Its intersection with a horizontal line at 1 is the instability window of $\omega_p$. For $\mathcal{L}(p) > 1$, subharmonic oscillation occurs.

**Example 3.** (*With phase margin of* $38.9°$ *based on the average model, the subharmonic oscillation still occurs.*) Consider a buck converter with the type-III compensator (27). Exactly the same parameters as in the technical document [18] are used: $f_s = 1/T = 300$ kHz, $L = 900$ nH, $C = 990$ $\mu$F, $R = 0.4$ $\Omega$, $R_c = 5$ m$\Omega$, $v_r = 3.3$ V, and $V_m = 1.5$ V. For the compensator, $K_c = 7.78 \times 10^4$, the zeros are $1/2\sqrt{LC} = 1.675 \times 10^4$, $1/\sqrt{LC} = 3.35 \times 10^4$, and the poles are $\omega_p = \omega_s/2 = 9.425 \times 10^5$, and $1/R_c C = 2.02 \times 10^5$.

Time-domain simulation (Fig. 21) shows that the subharmonic oscillation occurs when $v_s = 16$ V ($D \approx 0.206$). This is also confirmed by the exact sampled-data analysis with a sampled-data pole at -1 when the subharmonic oscillation occurs. Based on the average model (with $T(j\omega)$ as in (28)), the loop gain $T(j\omega)$ shown in Fig. 22 has a phase margin of $38.9°$ for $v_s = 16$. The frequency response also shows an infinite gain margin because the phase never reaches -180°, which means that no matter how much $v_s$ increases (to increase the loop gain), the converter is expected to be stable based on the average model. However, the subharmonic oscillation still occurs when $v_s = 16$.

With $D = 0.2$, the prediction of the critical $v_s$ by (30) is 17.1, close to the simulation result and the sampled-data analysis. As discussed above, the small error is due to the approximation of $T(s)$ in (29). □

Since the VMC buck converter with a type-III compensator is of case $\mathcal{C}_5$, same as the ACMC buck converter. The VMC buck converter with a type-III compensator also has an instability window of $\omega_p$ as shown in the next example.

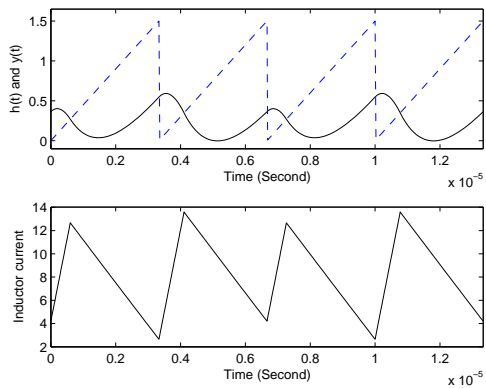

Figure 21.   Signal waveforms showing the subharmonic oscillation.

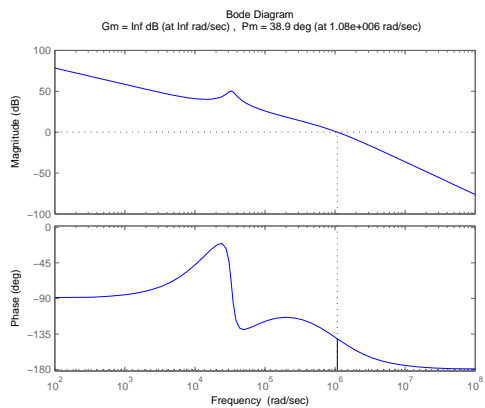

Figure 22.   Loop gain frequency response shows a phase margin of $38.9°$ and an infinite gain margin, but the subharmonic oscillation still occurs.

**Example 4.**   (*Unstable window of $\omega_p$, unrelated to the ripple amplitude of $y^0(t)$.*) Consider again Example 3 with $v_s = 16$. Vary $\omega_p$ from $0.1\omega_s$ to $0.6\omega_s$. An instability window of $\omega_p \in (0.23, 0.5)\omega_s$ is found. Similar to Example 2, the value of $\omega_p$ adjusts the ripple amplitude of $y^0(t)$. A larger $\omega_p$ leads to a larger ripple of $y^0(t)$. In [6], it is hypothesized that the ripple amplitude of $y^0(t)$ is related to subharmonic oscillation. The following simulation shows that the ripple amplitude of $y^0(t)$ is *unrelated* to subharmonic oscillation even for VMC.

**Time-domain simulation.** For $\omega_p = 0.2\omega_s$, the ripple amplitude of $y^0(t)$ is small, and the converter is stable (Fig. 23). For $\omega_p = 0.24\omega_s$, the ripple amplitude of $y^0(t)$ is larger, and the converter is unstable with subharmonic oscillation (Fig. 24). For $\omega_p = 0.6\omega_s$, the ripple amplitude of $y^0(t)$ (not shown) is even larger, but the converter is stable (Fig. 25). Comparing Figs. 23-25, the ripple amplitude of $y^0(t)$ is unrelated to subharmonic oscillation. This also shows a counter-example for the hypothesis proposed in [6] that the ripple amplitude of $y^0(t)$ is related to the subharmonic oscillation.

**Independent sampled-data analysis.** The instability window of $\omega_p$ is also confirmed by the sampled-data pole trajectories. The sampled-data pole trajectories for $0.1\omega_s < \omega_p < 0.6\omega_s$ are shown in Fig. 26. There are five poles. Three poles are fixed around $0.9485$, $0.8853$, and $0.51$. A pole leaves the unit circle through -1 when $\omega_p = 0.23\omega_s$, and enters the unit circle when $\omega_p = 0.5\omega_s$. This explains exactly the instability window of $\omega_p$.

**Accurate prediction by the L-plot.** The L-plot (where $T(s)$ is based on (29)) is shown in Fig. 27 which indicates an instability window of $\omega_p \in [0.23, 0.47]\omega_s$. As discussed above, the small error is due to the approximation of $T(s)$ in (29). The instability window of $p$ predicted by (26) is $[0.17, 0.58]$, which is also a close estimation.

In this example, the existence of instability window of $\omega_p$ is also verified by time-domain simulation, pole trajectories, and agreed closely with the derived critical condition.                                    □

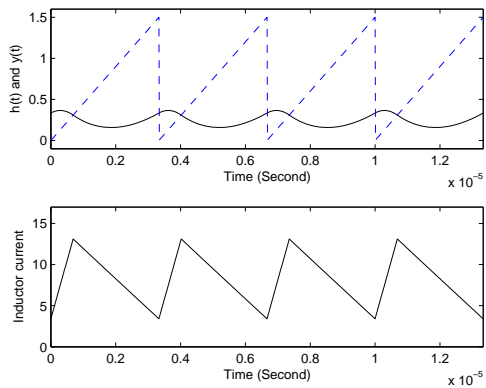

Figure 23. The converter is stable, $\omega_p = 0.2\omega_s$.

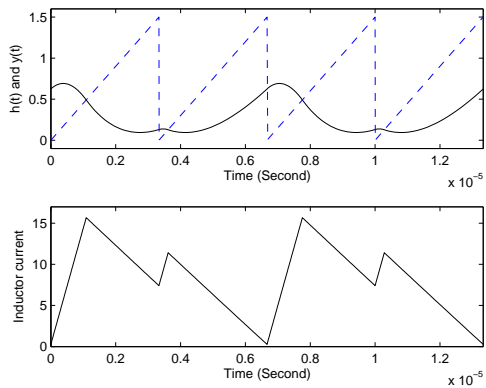

Figure 24. Subharmonic oscillation, $\omega_p = 0.24\omega_s$.

## X. CONCLUSION

A general critical condition (4) of subharmonic oscillation in terms of the loop gain is derived. It is applicable to a *general* nonlinear switching system represented in Fig. 5. Therefore, it is also applicable to other similar nonlinear control systems. Many *closed-form* critical conditions for various control schemes in terms of *converter parameters* are also derived. They are summarized in Table II. The effects of different parameters (such as $v_s$, $R$, $R_c$, and the ramp slope $m_a$) on the subharmonic oscillation can be clearly seen.

The questions asked in the Introduction are answered:

1) Those previously known critical conditions, such as (9) for CMC, (12) for V$^2$ control, and (16) for VMC with zero ESR, become special cases in the generalized framework.

2) The instability window of pole can be explained by case $\mathcal{C}_5$. A system of the case $\mathcal{C}_5$, such as ACMC or VMC with a type-III compensator, will have an instability window of pole.

3) The hypothesis [6] that the signal *ripple amplitude* can predict the occurrence of subharmonic oscillation is applicable for a converter with zero ESR (case $\mathcal{C}_6$). It is not applicable to a converter with a large pole (case $\mathcal{C}_5$). For example, in ACMC or VMC with a type-III compensator, the instability window of pole indicates that the signal ripple amplitude is not an accurate index to predict subharmonic oscillation. Also note that the ripple amplitude, either $\Delta v_o$ or $\Delta i_L$ for $R_c = 0$, has a term of $D(1 - D)$ which is symmetric with respect to $D = 1/2$. For a ripple index to be valid, the critical condition also needs to be symmetric with respect to $D = 1/2$. Among the cases in Table I, only the critical condition for $\mathcal{C}_6$ is symmetric with respect to $D = 1/2$. Therefore, the ripple index has very limited applications.

4) A typical critical condition is a weighted combination of three terms: $\alpha_0(D) = \pi(2D - 1)$, $\alpha_1(D) = \pi^2(2D^2 - 2D + 1)$, and a correction term $c(D, p)$. For example, both V$^2$ control and PVMC have the same form of critical condition. Also, both PVMC and CMC have a term $\alpha_0(D) = \pi(2D - 1)$. For a compensator with a pole smaller than one tenth of the switching frequency, the correction term can be ignored.

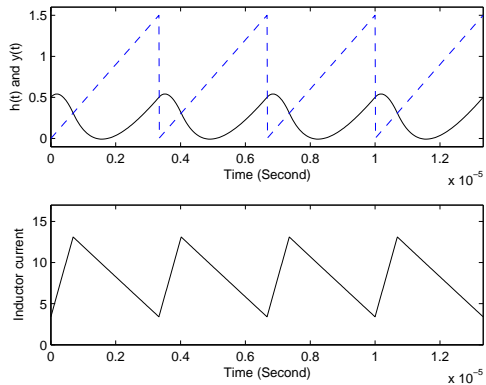

Figure 25. The converter is *stable* with a *larger* ripple of $y^0(t)$, $\omega_p = 0.6\omega_s$.

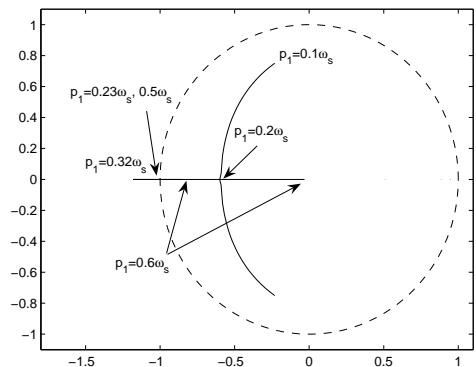

Figure 26. Sampled-data pole trajectories for $0.1\omega_s < \omega_p < 0.6\omega_s$.

5) Given an *arbitrary* control scheme, a *systematic* procedure is proposed to derive the critical condition for that control scheme. First, approximate the loop gain $T(s)$ at high frequency (higher than $\omega_s/2$). Then, from Table I, one can readily obtain the critical condition in terms of converter parameters.

Given the closed-form condition in terms of converter parameters, one knows the quantitative effect of each parameter on the subharmonic oscillation. One can make the L-plot (such as Figs. 20 and 27) as a function of a parameter of interest, its intersection with a horizontal line at 1 determines the stable operating range of that parameter. Also, given parameters $\mathsf{p}_1$ and $\mathsf{p}_2$ for example, based on the closed-form condition, one can determine the instability *boundary* in the parameter space $(\mathsf{p}_1, \mathsf{p}_2)$ as shown in Fig. 1, or from the contour plot (such as Fig. 15). Based on these plots, one knows how to choose proper parameter values to avoid the subharmonic oscillation.

This paper focuses on the converter operated at a fixed switching frequency. Similar analysis can be applied to the converter with variable frequency control, such as constant on-time control. The F-transforms in Table I still apply, but with a different $\alpha(D, p)$. The results are reported separately.

## REFERENCES

[1] R. W. Erickson and D. Maksimovic, *Fundamentals of Power Electronics*, 2nd ed. Berlin, Germany: Springer, 2001.

[2] D. C. Hamill, J. H. B. Deane, and J. Jefferies, "Modeling of chaotic DC-DC converters by iterated nonlinear mappings," *IEEE Trans. Power Electron.*, vol. 7, no. 1, pp. 25–36, 1992.

[3] R. B. Ridley, "A new, continuous-time model for current-mode control," *IEEE Trans. Power Electron.*, vol. 6, no. 2, pp. 271–280, 1991.

[4] F. D. Tan and R. D. Middlebrook, "A unified model for current-programmed converters," *IEEE Trans. Power Electron.*, vol. 10, no. 4, pp. 397–408, 1995.

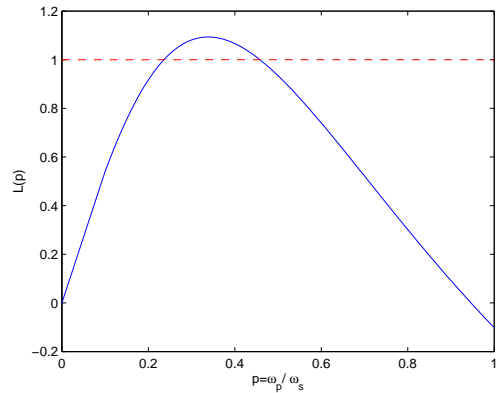

Figure 27.   L-plot as a function of $p$. Its intersection with a horizontal line at 1 is the instability window of $\omega_p$. For $\mathcal{L}(p) > 1$, subharmonic oscillation occurs.

Table II
CRITICAL CONDITIONS FOR DIFFERENT CONTROL SCHEMES.

| | |
|---|---|
| $\mathcal{C}_2$ | CMC, $\frac{v_s}{L}(D - \frac{1}{2}) = m_a$ |
| $\mathcal{C}_7$ | PVMC, $R_c \neq 0$, $\frac{v_s k_p \rho T^2}{4 V_m L C} \left[ \frac{2 R_c C}{T} (2D - 1) + (2D^2 - 2D + 1) \right] = 1$ |
| $\mathcal{C}_7$ | $V^2$ Control, same as PVMC above |
| $\mathcal{C}_6$ | PVMC with $R_c = 0$, $\frac{v_s k_p T^2}{4 V_m L C} (2D^2 - 2D + 1) = 1$ |
| $\mathcal{C}_5$ | PVMC with $R_c = 0$ and Small $R$, $v_s = \frac{L V_m \omega_s}{R k_p (\alpha_0(D) - \alpha(D,p))}$ |
| $\mathcal{C}_3$ | PVMC $RL$ circuit, $v_s k_p p \alpha(D,p) = V_m$ |
| $\mathcal{C}_5$ | ACMC with type-II compensator, $v_s = \frac{V_m z_c L \omega_s}{R_s K_c (\alpha_0(D) - \alpha(D,p))}$ |
| $\mathcal{C}_5$ | VMC with type-III compensator, $v_s = \frac{V_m z_s \omega_s}{K_c \rho (\alpha_0(D) - \alpha(D,p))}$ |

[5] C.-C. Fang, "Sampled data poles, zeros, and modeling for current mode control," *International Journal of Circuit Theory and Applications*, 2011, accepted and published online, DOI: 10.1002/cta.790.

[6] A. El Aroudi, E. Rodriguez, R. Leyva, and E. Alarcon, "A design-oriented combined approach for bifurcation prediction in switched-mode power converters," *IEEE Transactions on Circuits and Systems II: Express Briefs*, vol. 57, no. 3, pp. 218–222, Mar. 2010.

[7] R. Redl and J. Sun, "Ripple-based control of switching regulators - an overview," *IEEE Trans. Power Electron.*, vol. 24, no. 12, pp. 2669–2680, Dec. 2009.

[8] C.-C. Fang, "Modeling and instability of average current control," in *EPE International Power Electronics And Motion Control Conference*, 2002, paper SSIN-03.

[9] J. Li and F. C. Lee, "Modeling of $V^2$ current-mode control," *IEEE Transactions on Circuits and Systems-I: Regular Papers*, vol. 57, no. 9, pp. 2552–2563, 2010.

[10] R. Genesio and A. Tesi, "Harmonic balance methods for the analysis of chaotic dynamics in nonlinear systems," *Automatica*, vol. 28, no. 3, pp. 531–548, 1992.

[11] A. Tesi, E. H. Abed, R. Genesio, and H. O. Wang, "Harmonic balance analysis of period-doubling bifurcations with implications for control of nonlinear dynamics," *Automatica*, vol. 32, no. 9, pp. 1255–1271, 1996.

[12] C.-C. Fang, "Sampled-data analysis and control of DC-DC switching converters," Ph.D. dissertation, Dept. of Elect. Eng., Univ. of Maryland, College Park, 1997, available: http://www.lib.umd.edu/drum/, also published by UMI Dissertation Publishing in 1997.

[13] C.-C. Fang and E. H. Abed, "Harmonic balance analysis and control of period doubling bifurcation in buck converters," in *Proc. IEEE ISCAS*, vol. 3, May 2001, pp. 209–212.

[14] ——, "Analysis and control of period-doubling bifurcation in buck converters using harmonic balance," *Latin American Applied Research: An International Journal*, pp. 149–156, 2001, Special theme issue: Bifurcation Control: Methodologies and Applications, In Honor of the 65th Birthday of Professor Leon O. Chua.

[15] M. K. Kazimierczuk, "Transfer function of current modulator in PWM converters with current-mode control," *IEEE Transactions on Circuits and Systems-I: Fundamental Theory and Applications*, vol. 47, no. 9, pp. 1407–1412, 2000.

[16] C. P. Basso, *Switch-Mode Power Supplies*.   McGraw-Hill, 2008.

[17] W. Tang, F. C. Lee, and R. B. Ridley, "Small-signal modeling of average current-mode control," *IEEE Trans. Power Electron.*, vol. 8, no. 2, pp. 112–119, 1993.

[18] D. Mattingly, "Designing stable compensation networks for single phase voltage mode buck regulators," Intersil Americas Inc., Tech. Rep., 2003, available: www.intersil.com/data/tb/tb417.pdf.